\documentstyle[aps,prl,epsf]{revtex}
\begin{document}
\title {Anomalous Scaling in Kolmogorov--1941 Turbulence}
\author{Victor S. L'vov$^{*}$ and Vladimir V. Lebedev$^{\dag}$ }
\address{ Department of Physics, Weizmann Institute
of Science, Rehovot, 76100, Israel \ \ \ \ {\rm and} \\
$^{*}$Institute of Automation and Electrometry, Acad. of Sci. of
Russia, 630090, Novosibirsk, Russia\\
$^{\dag}$Landau Inst. for Theor. Phys., Acad. of Sci. of Russia,
117940, GSP-1, Moscow, Kosygina 2, Russia}
\maketitle
\begin{abstract}
We show that the Kolmogorov-1941 picture of fully developed hydrodynamic
turbulence (with the scaling of the structure functions $S_n(R) \propto
R^{n/3}$) necessarily leads to an anomalous scaling for correlation
functions which include the rate of energy dissipation $\varepsilon
(t,{\bf r})$, these correlation functions being described by an
independent index. The mechanism for anomalous scaling, suggested on the
basis of the Navier-Stokes equation, is the multi-step interaction of
eddies from the inertial interval with eddies at the viscous scale via a
set of eddies of intermediate scales.
\end{abstract}
\pacs{PACS numbers 47.10,47.25C}

% \noindent
{\bf Introduction.}\ \
The modern theory of hydrodynamic turbulence originates from the
Kolmogorov's concept of turbulence (hereafter K41) as a step-by-step
cascade of energy over scales \cite{K941,O941}. According to K41 in the
inertial range of scales between the outer (pumping) scale $L$ and inner
(viscous) scale $\eta =(\nu^3/\bar \varepsilon )^{1/4}$  the only relevant
parameter is $\bar\varepsilon$, the mean value of the energy dissipation
rate
\begin{equation}
\varepsilon ( {\bf r}) =\case{1}{2}\nu
\big[\nabla_\alpha v_\beta ( {\bf r})
+ \nabla_\beta v_\alpha ( {\bf r})\big]^2\ .
\label{J01} \end{equation}
For the structure functions with $L\gg R \gg \eta$ K41 dimensional
reasoning yields
\begin{equation}
S_n(R)\equiv\langle \mid  {\bf v}( {\bf r})
- {\bf v}(  {\bf r}+{\bf R}) \mid^n \rangle \sim
(\bar\varepsilon R)^{n/3}  \ .
\label{J02} \end{equation}
The widely spread belief is that intermittency (fluctuations of
$\varepsilon$) modifies (\ref{J02}) leading to $S_n \propto R^{\zeta_n}$,
$\zeta_n\ne n/3$ \cite{MY75}. To characterize intermittency
one introduces the correlation function of $\varepsilon$-fluctuations
$\tilde{\varepsilon}( {\bf r}) = \varepsilon( {\bf r})-\bar{\varepsilon}$
which is believed to have a scaling form in the inertial range
\begin{equation}
K_{\varepsilon \varepsilon }(R)= \langle
\tilde{\varepsilon} ({\bf r})\tilde{\varepsilon} ( {\bf r}+{\bf R})
\rangle \propto R^{-\mu} \,,
\label{J03}
\end{equation}
with a phenomenological constant $\mu$ \cite{MY75}. Numerous measurements
to determine the exponent $\mu$ (see e.g. \cite{MS93}) have resulted in a
set of values ranging from $0.15$ to $0.4$. Various phenomenological
models of intermittency give different relationships between $\zeta_n$ and
$\mu$; in some models \cite{K962,NS64,FSN4} the ``bridge'' $\zeta_6=2-\mu$
arises. However this constraint is questionable \cite{MS91,F993} and
really one cannot solidly relate $\zeta_n$ and $\mu$ using only
phenomenology.  A diagrammatic analysis based on the Navier-Stokes (NS)
equation \cite{BL87,L991,LL93} shows that in the limit ${\rm Re}\to\infty$
fluctuations of $\varepsilon$ would not change (\ref{J02}). Thus we accept
the K41 scaling and will try to examine the correlation function
(\ref{J03}) at this assumption.

The naive estimation of the correlation function
$K_{\varepsilon\varepsilon}(R)$ in the framework of K41 gives \cite{G962}
\begin{equation}
K_{\varepsilon \varepsilon}^{^{\rm G}}(R) \sim \nu^2
\big({d^2 S_2(R)/ d R^2}\big)^2
\sim \bar \varepsilon ^2 ({\eta/R})^{8/3}
\label{J04}
\end{equation}
in obvious contradiction with the experiment. The conventional way out (see
e.g.\cite{YZ93}) is that one should take into account the nonlinear
interaction by replacing in (\ref{J04}) the molecular viscosity $\nu$ with the
turbulent viscosity $\nu_{_{\rm T}}(R)$. In K41 $\nu_{_{\rm T}}(R)\propto
R^{4/3}$; this yields $\mu=0$. However this procedure is not justified. In
this Letter we develop an analytical theory of $\varepsilon$--$\varepsilon$
correlations based on the diagrammatic approach which predicts a strong
renormalization of the naive estimation (\ref{J04}). Let us stress that this
renormalization has noting in the replacement $\nu\rightarrow\nu_{_{\rm
T}}(R)$. In our theory $\mu$ remains an independent parameter which we cannot
yet calculate explicitly. In order to elucidate the physical basis of the
involved diagrammatic expansion we first describe our findings using the
popular handwaving language of cascades, eddies and their interactions before
turning to an overview of the cumbersome technicalities of the presented
theory.

{\bf Telescopic Multi-Step Eddy Interaction.}\ \
For our analysis it is useful to present the velocity field as a
 superposition of eddies of various sizes $x$. In K41 the velocity
gradient scales like $x^{-2/3}$, therefore the main contribution to
$\tilde{\varepsilon}$ itself is expected to come from eddies of the size
of the order of $\eta$ while the main contribution to the correlation on
the distance $R$ is expected from the eddies which bridge the gap between
${\bf r}$ and ${\bf r}+{\bf R}$. This means that scales larger than $R$
can be neglected in our consideration. We will show that the direct effect
of the $R$-eddies (of size $R$) produces the naive estimation (\ref{J04})
while the indirect influence (via the eddies of intermediate scales) leads
to the strong renormalization of the index $\mu$.

To this end we portion the range from $\eta$ to $R$ into $N$ subintervals
$\eta<x_1<x_2<\dots<x_N$ where $R$ is well within the $N$-th subinterval.
This can be thought as a partition of $k$-space into shells separated by
wavevectors $k_n=2\pi/x_n$. The velocity of $n$-eddies ${\bf v}_n({\bf
x})$ is the sum of Fourier harmonics with $k$ between $k_{n-1}$ and $k_n$.
To find the correlation function $K_{\varepsilon\varepsilon}$ we should
perform averaging over the statistics of eddies in all subintervals.  The
crucial point is that the statistics of eddies in different subintervals
is not independent. Namely we should take into account the effect of
changing statistics of small eddies in the strain field of larger eddies.

The correlation between fluctuations on different scales can be described
by a conditional probability density. We will designate as $P_n$ the
probability density to find near $\bf r$ the velocity gradient
$\nabla_\alpha{\bf v}$ of $n$-eddies.  The quantity $P_n$ is a function of
$\nabla_\alpha{\bf v}_{>n}$ which is the sum of gradients of the velocity
fields of larger eddies:  ${\bf v}_{>n}={\bf v}_{n+1}({\bf r})+{\bf
v}_{n+2}({\bf r})+\dots$.  K41 scaling implies that the velocity gradients
of larger scales are comparatively small. Hence, we will make use of the
expansion of $P_n$ for small large-scale gradients:
\begin{equation}
P_n = P_n^{(0)} +P_n^{(2)}\cdot
(\nabla_\alpha {\bf v}_{>n})^2 +\dots
\quad , \label{J95} \end{equation}
where $P_n^{(0)}$, $P_n^{(2)}$ are functions of $\nabla_\alpha {\bf v}_n$
only. The linear term of the expansion over $\nabla_\alpha {\bf v}_{>n}$
is absent in (\ref{J95}) because of the incompressibility condition
${\bbox\nabla} \cdot {\bf v}_{>n}=0$.

We illustrate the consequences of such correlations by considering the
hypothetical case where only fluctuations of $N$- and $n$-eddies ($n<N$)
are excited. Then designating by halfsquare brackets the averaging over
the statistics of corresponding eddies we find
\begin{equation}
K_{\varepsilon \varepsilon }(R)
=\big \lfloor \lfloor \tilde{\varepsilon}( {\bf r})
\tilde{\varepsilon}( {\bf r}+{\bf R})
\rfloor_n\big\rfloor_{_N}
=\big \lfloor \lfloor
\tilde{\varepsilon}( {\bf r}) \rfloor_n
\lfloor \tilde{\varepsilon}( {\bf r}+{\bf R})
\rfloor_n\big\rfloor_{_N}
\quad , \label{J05} \end{equation}
because the correlation length under the conditional $n$-average is much
shorter than $R$. Starting from (\ref{J01}) and using the expansion of the
conditional probability we obtain
\begin{equation}
\lfloor \tilde\varepsilon({\bf r})\rfloor_n =
\tilde\varepsilon_{_N}({\bf r})+
B_n\tilde\varepsilon_{_N}({\bf r}) \quad ,
\label{J06} \end{equation}
where $\tilde\varepsilon_{_N}({\bf r})=
\nu\big(\nabla v_{_N}( {\bf r})\big)^2
-\nu\lfloor\big(\nabla v_{_N}( {\bf r})\big)^2 \rfloor_{_N}$.
The first term on the RHS of (\ref{J06}) is the direct contribution of the
$N$-eddies to $\lfloor\tilde{\varepsilon}({\bf r})\rfloor_n$, the second
term derives immediately from the expansion of $P_n$.  The largest
contribution to $\lfloor\varepsilon({\bf r})\rfloor_n$ comes from the
$n$-eddies themselves, this is $\nu \lfloor [\nabla v_n( {\bf
r})]^2\rfloor_n$. However this contribution is independent of time and
space coordinates and is canceled by subtracting $\bar \varepsilon$.

The coefficient ${B}_n$ in (\ref{J06}) can be estimated applying physical
reasoning based on NS equations. The dimensionless parameter describing
the relative change of velocity ${\bf v}_n( {\bf r})$ with varying
$\nabla_\alpha{\bf v}_{_N}$ is $\tau_n\nabla {v}_{_N}({\bf r})$, where
$\tau_n$ is the life time of $n$-eddies, hence $B_n \sim \tau_n^2 \lfloor
|\nabla v_n|^2 \rfloor_n$. Since in K41 the life time $\tau_n$ coincides
with the turnover time of the $n$-eddies we conclude $ B_n  \sim r_n^
{0}$, which means that $B_n$ is independent of the scale $r_n$.  This
statement is of crucial importance for understanding the origin of the
anomalous scaling:  all scales in the interval from $R$ to $\eta$
contribute equally to $K_{\varepsilon\varepsilon}$. Therefore we have to
take into account not only contributions from two scales, as we did up to
now, but contributions of all the scales from the above interval.
\begin{figure}
\epsfxsize=8.6truecm
\epsfbox{fig.eps}
\end{figure}
\vspace{-4.8cm}
\hspace{10cm}
\parbox[b]{7cm}{FIG. 1
 Telescopic eddy interaction of three groups of eddies of scales
       $R$, $x_n$, and $x_m$.  Ellipses 1 and 2 show eddies of the
       scale $\eta$, separated by the distance $R$. For contributions of
       various interactions in the ``telescope'', see Eq.
       (\protect\ref{J09}).
\vspace{2cm} }

Now let us consider the case of three groups of eddies of scales
$x_{_N}> x_n> x_m$ which will indicate the general form of (\ref{J06})
when all eddies are present. These eddies are depicted in Fig. 1.
%\ref{fig:fig}.
Instead of the contribution $\lfloor\tilde\varepsilon({\bf r})\rfloor_n$
in (\ref{J05}) one has now:
\begin{equation}
\lfloor\tilde\varepsilon( {\bf r})\rfloor_{m,n}
=\tilde\varepsilon_{_N}({\bf r})
\big(1+B_n+B_m+B_n\,B_m \big)
\quad . \label{J09} \end{equation}
The first term on the RHS of (\ref{J09}) describe the direct contribution
of the $R$-eddies, the second and the third terms are associated with the
influence of $R$-eddies on the $n$- and $m$-eddies, respectively.  The
last term ($\propto B_n\,B_m $) is due to the indirect effect of the
largest scale, the $R$-eddies, on the smallest scale $x_m$ via the
intermediate $n$-eddies. To obtain this term one has to repeat twice the
above expansion. The RHS of (\ref{J09}) is proportional to
$(1+B_n)(1+B_m)$.  Thus it is plausible that one obtains in the general
case
\begin{equation}
\lfloor\tilde\varepsilon( {\bf r})\rfloor
=\tilde\varepsilon_{_N}({\bf r})
\prod_{n=1}^N \big(1+B_n\big) \,,
\label{J10}
\end{equation}
which should be substituted into (\ref{J05}) instead of
$\lfloor\tilde{\varepsilon}({\bf r})\rfloor_n$.  Note that the
independence of the $B_n$ of $n$ as pointed out above will only hold when
the width $\Delta k_n$ of the shells scales in the same way as the $k_n$
themselves. We choose $k_{n+1}/k_n=\Lambda$ so that neighboring shells may
be considered as almost statistically independent. Such $\Lambda>1$ does
exist because of the locality of energy transfer via the scales
\cite{BL87}.  Then one can rewrite (\ref{J10}) as
\begin{equation}
\lfloor \tilde\varepsilon( {\bf r})\rfloor
=\tilde\varepsilon_{_N}({\bf r})
\big(1+B_n\big)^N
\sim \tilde\varepsilon_{_N}({\bf r}) \big(R/\eta\big)^\Delta
\label{J11}
\end{equation}
where $N=$log$_\Lambda(R/\eta)$ and $\Delta=\ln(1+B)/\ln (\Lambda)$.
Following the terminology accepted in the theory of phase transitions
one can call the exponent $\Delta$ an {\it anomalous dimension} of
$\varepsilon$. From (\ref{J05},\ref{J11}) one obtains finally
\begin{equation}
K_{\varepsilon\varepsilon}(R)\sim
\nu^2\big(R/\eta\big)^{2\Delta}
\lfloor\tilde\varepsilon_{_N}({\bf r})
\tilde\varepsilon_{_N}({\bf r}+{\bf R})\rfloor_{_N}
\sim \bar\varepsilon^2 (\eta/R)^\mu
\label{J12}
\end{equation}
with $\mu=8/3-2\Delta$ since the average over $N$-eddies can be estimated
naively as in (\ref{J04}). To explain the experimental value of $\mu<0.5$
the value of $\Delta$ should be near $1$ which corresponds to the strong
renormalization of the naive value.

{\bf Diagrammatic equations.}\ \
We will use the Wyld diagrammatic technique \cite{W941} with the
Belinicher-L'vov resummation\cite{BL87} (see also \cite{L991}). This
enables us to represent any correlation function characterizing a
turbulent flow as a series over the different-time correlation function
$F_{\alpha \beta }$ of the quasi-Lagrangian (qL) velocity differences
${\bf w}(t,{\bf r})={\bf v}(t,{\bf r})- {\bf v}(t,0)$ and corresponding
Greens' function $G_{\alpha \beta}$
\begin{equation}
F_{\alpha \beta }(t,{\bf r}_1,{\bf r}_{_2})=
\langle w_\alpha(t,{\bf r}_1)
w_\beta(0,{\bf r}_{_2}) \rangle \,, \quad
G_{\alpha \beta }(t ,{\bf r}_1,{\bf r}_{_2})=
-i\delta \langle  w_\alpha (t,{\bf r}_1) \rangle
/ \delta \langle f_\beta (0,{\bf r}_2) \rangle
\quad . \label{J17} \end{equation}
The qL velocity ${\bf v}(t,{\bf r})$ is related to the Eulerian velocity
${\bf u}(t,{\bf r})$ according to\cite{BL87} ${\bf u}(t,{\bf r})={\bf
v}\left(t,{\bf r}-\int^t {\bf v}(\tau,0)d\,\tau\right)$.  The $G$-function
is defined here as the susceptibility determining the average $\langle
{\bf w} \rangle$ which appears if the external force ${\bf f}$ in the RHS
of NS equation has a nonzero average $\langle{\bf f}\rangle$.  The above
formulation is an order-by-order Galilei invariant and has no infrared
divergences of the integrals in all diagrams.  Therefore the external
scale of the turbulence is absent in diagrammatic equations. We showed
that K41 is the only scale-invariant solution of these equations
\cite{LL93}. The correlation functions (\ref{J17}) may be evaluated as
\begin{equation} G(t, {\bf r}_1, {\bf r}_{_2})  \propto R^{-3} \,,
\quad F(t, {\bf r}_1, {\bf r}_{_2}) \propto R^{2/3} \,,
\label{J19}
\end{equation}
where $R$ is the characteristic scale.

The correlation function $K_{\varepsilon\varepsilon}$ is represented by an
infinite series of diagrams.  In the spirit of the Keldysh diagrammatic
technique \cite{K964} (the Wyld technique is the classical limit of the
Keldysh one) we will utilize only one type of line designating both $F$-
and $G$-functions. Interaction vertices on diagrams are determined by the
nonlinear term in the qL version of NS equation\cite{BL87,L991}. Diagrams
representing contributions to $K_{\varepsilon\varepsilon}$ are depicted in
Fig. 2.  A circle on these diagrams corresponds to (\ref{J01}). A
rectangle on these diagrams represents the sum of diagrams which cannot be
cut into two parts along two horizontal lines. The first contributions to
the rectangle are determined simply by the $F$ or $G$-line. An analysis of
higher order contributions analogous to that produced in \cite{BL87,L991}
shows that the dimension estimate of the rectangle given by the first
contributions is reproduced for the entire sum of diagrams and is
consequently determined by (\ref{J19}).
\vspace{-.5cm}
\begin{figure}
\setlength{\unitlength}{.2cm}
\begin{picture}(58,14)(0,-4)
\thicklines
\put(2,4){\circle{1}}
\put(8,4){\circle{1}}
\put(5,4.5){\oval(6,5)[t]}
\put(5,3.5){\oval(6,5)[b]}
\put(10,3){\line(0,1){2}}
\put(9,4){\line(1,0){2}}
\put(12,4){\circle{1}}
\put(20,4){\circle{1}}
\put(16,4.5){\oval(8,5)[t]}
\put(16,3.5){\oval(8,5)[b]}
\put(15.5,1){\line(0,1){6}}
\put(16.5,1){\line(0,1){6}}
\put(22,3){\line(0,1){2}}
\put(21,4){\line(1,0){2}}
\put(24,4){\circle{1}}
\put(36,4){\circle{1}}
\put(30,4.5){\oval(12,5)[t]}
\put(30,3.5){\oval(12,5)[b]}
\put(27.5,1){\line(0,1){6}}
\put(28.5,1){\line(0,1){6}}
\put(31.5,1){\line(0,1){6}}
\put(32.5,1){\line(0,1){6}}
\put(38,3){\line(0,1){2}}
\put(37,4){\line(1,0){2}}
\put(40,4){\circle*{0.4}}
\put(41,4){\circle*{0.4}}
\put(42,4){\circle*{0.4}}
\put(43,3.5){\line(1,0){2}}
\put(43,4.5){\line(1,0){2}}
\put(47,4){\circle*{2}}
\put(55,4){\circle*{2}}
\put(51,4){\oval(8,6)}
\thinlines
\put(2,-3){\line(0,1){2.5}}
\put(8,-3){\line(0,1){2.5}}
\put(16,-3){\line(0,1){2.5}}
\put(20,-3){\line(0,1){2.5}}
\put(28,-3){\line(0,1){2.5}}
\put(32,-3){\line(0,1){2.5}}
\put(36,-3){\line(0,1){2.5}}
\put(4,-2.5){$R$}
\put(17,-2.5){$x$}
\put(33,-2.5){$x$}
\put(29,-2.5){$x_1$}
\put(50, -2.5){FIG. 2. The sum of the diagrams for
$K_{\varepsilon\varepsilon}(R)$.}
\end{picture}
%\caption{The sum of the diagrams for $K_{\varepsilon\varepsilon}(R)$.}
%\label{fig:fig1}
\end{figure}
%\vspace{1cm}
Based on this result we conclude that the expression corresponding to the
right loop in the second diagram in Fig. 3 contains a logarithmic factor
originating from scales $R\gg x \gg \eta$.  The third diagram in Fig. 2
will give the second power of the logarithm originating from the region of
scales $R\gg x\gg x_1 \gg \eta$ (we would like to stress that the region
$R\gg x_1\gg x \gg \eta$ does not give the second power of the logarithm).
Therefore the problem of summing the main logarithmic terms appears. To
solve it let us take the cut corresponding to the largest separation of
the order of $R$. Then the parts of the diagrams on the left and on the
right of this cut will give after summation a three-leg object which we
depict as a disk. As a result we come to the representation for
$K_{\varepsilon\varepsilon}(R)$ depicted in Fig. 2.

The three-leg object differs from the bare object by a factor ${\cal N}(R)$.
Resumming the ``ladder'' sequence of diagrams we find for ${\cal N}(x)$ the
integral equation depicted in Fig. 3. In analytical form it is
\begin{equation}
{\cal N}(x) = 1 + \int_\eta^x
dy \, A(x,y) {\cal N} (y)
\quad , \label{J20} \end{equation}
where $A(x,y)$ is a uniform function with the index $-1$. This allows
a power-like solution
\begin{equation}
{\cal N}(x)  \sim (x/\eta)^\Delta  \,,
\label{J21}
\end{equation}
where $\Delta$ is the index coinciding with the ones introduced in the
preceding Section. Let us stress that $\Delta$ cannot be zero because of
the logarithmic divergence at small $y$ noted above. The problem is that
in the region $x\gg y$ the kernel $A(x,y)\propto y^{-1}$.  This law is the
consequence of the asymptotic behavior of the qL vertex which is
proportional to the smallest wavenumber \cite{BL87,L991}.

\begin{figure}
\setlength{\unitlength}{.2cm}
\begin{picture}(34,8)
\thicklines
\put(3,4){\circle*{2}}
\put(3,5){\line(0,1){1}}
\put(3,2){\line(0,1){1}}
\put(6,3.5){\line(1,0){2}}
\put(6,4.5){\line(1,0){2}}
\put(11,4){\circle{2}}
\put(11,5){\line(0,1){1}}
\put(11,2){\line(0,1){1}}
\put(15,3){\line(0,1){2}}
\put(14,4){\line(1,0){2}}
\put(20,1){\line(0,1){6}}
\put(21,1){\line(0,1){6}}
\put(31,4){\circle*{2}}
\put(19,5){\oval(24,4)[tr]}
\put(19,3){\oval(24,4)[br]}
\put(40,4){FIG. 3.
The equation for the renormalization factor ${\cal N}(x)$.}
\end{picture}
%\caption{The equation for the renormalization factor ${\cal N}(x)$.}
%\label{fig:fig2}
\end{figure}

Now we can proceed to the analysis of the scaling behavior of
$K_{\varepsilon\varepsilon}$. To do this we can use the diagrammatic
representation for $K_{\varepsilon\varepsilon}$ given in Fig. 2 (the last
term). If one takes the bare value ${\cal N}=1 $ the last diagram in Fig.
2 will reproduce the expression (\ref{J04}). If one takes the dressed
functions ${\cal N} $ we conclude that the last diagram in Fig. 2 gives
the contribution $\propto R^{2\Delta-8/3}$.  The above diagrammatic
analysis obviously leads to the same results (\ref{J12}) as the
semi-qualitative analysis given in the previous Section.

{\bf Discussion.}\ \
We have demonstrated that assuming K41 scaling (\ref{J02}) one finds an
essential renormalization of the naive K41 value $\mu=8/3$ (the
experimental value is $\mu<0.5$) due to the multi-step eddy interaction.
Note that such renormalization mechanism works for any scaling of the
velocity correlation functions with $\zeta_n\propto n$. Although we cannot
find the renormalization explicitly we see that $\mu$ is determined by
complex integrals of correlation functions.  Therefore one could not
expect a simple relation between $\mu$ and $\zeta_n$ like the ``bridge''
$\zeta_6-\mu$. Note also that the same renormalized exponent $\mu$ should
determine the scaling of other objects e.g. the asymptotic behavior of the
four-point correlation function of velocities \cite{LL94}.  Let us stress
that in our theory both the structure functions $S_n$ and the correlation
functions of $\tilde\varepsilon$ do not depend of the outer scale of
turbulence $L$ at Re$\to\infty$, whereas experiment seems to demonstrate
$L$-dependence of different correlation functions. This could be related
to a transient regime at finite Re. In any case (as the physics of second
order phase transitions shows) in order to compare adequately experiment
and theory one needs suitable interpolation formulae. Obtaining such
formulae is a separate problem which is out of the scope of the Letter.

{\bf Acknowledgments.}\ \
Numerous discussion with Robert Kraichnan, Itamar Procaccia and Stefan
Thomae were very illuminative for us. We are grateful to Adrienne Fairhall
for help. This work has been supported by the Minerva Center for Nonlinear
Physics of Complex Systems and the Landau--Weizmann Program.

\end{document}